\documentclass[aps,prd,10pt,twocolumn,floatfix,
  noshowpacs,preprintnumbers,superscriptaddress,nofootinbib]{revtex4-1}
\usepackage[utf8]{inputenc}  
\usepackage{amsmath}
\usepackage{amssymb} 

\usepackage{enumerate}
\usepackage{amsmath}  
\usepackage{tikz}
\usepackage{tipa}
\usepackage{CJKutf8}
\usepackage{feynmf}
\usepackage{slashed}
\usepackage{braket}
\usepackage[toc,page]{appendix}
\usepackage{url}
\usepackage{natbib}
\usepackage{graphicx}
\usepackage[pdftex,bookmarks,linktocpage,pdfpagelabels,plainpages=false,hyperfigures,linkcolor=blue,citecolor=blue]{hyperref} 
\hypersetup{colorlinks=true}

\usepackage{mathrsfs,amssymb}  
\usepackage{cancel}
\usepackage[normalem]{ulem}
\usepackage{array}
\usepackage{booktabs}
\usepackage{verbatim}

\begin{document}
\begin{flushright}
MI-HET-782
\end{flushright}

\author{Wooyoung~Jang}
\affiliation{Department of Physics, University of Texas, Arlington, TX 76019, USA}

\author{Doojin~Kim}
\affiliation{Mitchell Institute for Fundamental Physics and Astronomy,
Department of Physics and Astronomy, Texas A\&M University, College Station, TX 77843, USA}

\author{Kyoungchul Kong}
\affiliation{Department of Physics and Astronomy, University of Kansas, Lawrence, KS 66045, USA}

\author{Youngjoon Kwon}
\affiliation{Department of Physics \& IPAP, Yonsei University, Seoul 03722, Republic of Korea}

\author{Jong-Chul Park}
\affiliation{Department of Physics and IQS, Chungnam National University, Daejeon 34134, Republic of Korea}

\author{Min Sang Ryu}
\affiliation{Department of Physics, University of Seoul, Seoul 02504, Republic of Korea}
\affiliation{Department of Physics, Kyungpook National University, Daegu 41566, Republic of Korea}

\author{Seodong Shin}
\affiliation{Department of Physics, Jeonbuk National University, Jeonju, Jeonbuk 54896, Republic of Korea}

\author{Richard G. Van de Water}
\affiliation{Los Alamos National Laboratory, Los Alamos, NM 87545, USA}

\author{Un-Ki Yang}
\affiliation{Department of Physics, Seoul National University, Seoul 08826, Republic of Korea}

\author{Jaehoon~Yu}
\affiliation{Department of Physics, University of Texas, Arlington, TX 76019, USA}

\title{Search Prospects for Axion-like Particles at \\ Rare Nuclear Isotope Accelerator Facilities}

\begin{abstract} 
We propose a novel experimental scheme, called DAMSA (Dump-produced Aboriginal Matter Searches at an Accelerator), for searching for dark-sector particles, using rare nuclear isotope accelerator facilities that provide high-flux proton beams to produce a large number of rare nuclear isotopes.
The high-intensity nature of their beams enables the investigation of dark-sector particles, including axion-like particles (ALPs) and dark photons. 
By contrast, their typical beam energies are not large enough to produce the backgrounds such as neutrinos resulting from secondary charged particles. 
The detector of DAMSA is then placed immediate downstream of the proton beam dump to maximize the prompt decay signals of dark-sector particles, which are often challenging to probe in other beam-dump-type experiments featuring a longer baseline, at the expense of an enormous amount of the beam-related neutron (BRN) background. We demonstrate that BRN can be significantly suppressed if the signal accompanies multiple, correlated visible particles in the final state.
As an example physics case, we consider ALPs interacting with the Standard Model photon and their diphoton decay signal at DAMSA implemented at a rare nuclear isotope facility similar to the Rare isotope Accelerator complex for ON-line experiment under construction in South Korea.  
We show that the close proximity of the detector to the ALP production dump makes it possible to probe a high-mass region of ALP parameter space that the existing experiments have never explored.
\end{abstract}

\maketitle

\noindent {\bf Introduction.} 
Dark matter or more generally dark-sector physics is a well-motivated paradigm for extensions of the Standard Model (SM), and a wide range of models have been proposed, predicting various new particles. 
Of them, new particles including axion-like particles (ALPs) and dark photon are receiving particular attention, as they often mediate the interactions between SM and dark-sector particles.
Their feebly interacting nature, however, makes it challenging for energy-frontier experiments to discover such mediators; it is more probable that the first discovery of them would be made in an experiment at the facilities with high-intensity beams, such as a beam-dump type experiment.

A number of ongoing beam-dump type experiments mostly focus on the decay of mediators to SM particles and have set stringent limits in their parameter space. 
By construction, once a mediator is produced in a target, it should be sufficiently long-lived to enter the detector before decaying to lighter particles. 
It, however, should then decay inside the detector to be detected.
The balance between its survival and decay governs the shape of the sensitivity reaches at a given experiment, highly depending on the details of the experimental configuration such as the detector geometric coverage, the distance between the production location and the detector, etc.
While higher beam intensity enables expanding the coverage of parameter space, the region associated with high-mass mediators with large coupling strengths is challenging to explore due to the prompt decay of the mediators; the access to the high-mass range of the parameter space is determined by the distance between the mediator production vertex and the detector.

In this regard, we point out that high-intensity proton beam facilities for rare nuclear isotope production (e.g., FRIB~\cite{Wei:2019oew}, ISOLDE~\cite{Catherall:2017kbr}, and RAON~\cite{Jeon:2014oja}) possess the possibility of placing a precision large-scale detector immediate downstream of the dump, greatly reducing the distance between the source and the detector. 
Since the available center-of-mass frame energy (e.g., $\sim 280$~MeV from a 600-MeV beam collision on a target) is barely above twice the pion mass, the flux of charged pions is significantly reduced and so is that of neutrinos from their decay, leaving the beam-related neutrons (BRN) to be the major background which can be significantly suppressed for the signals with features.

Exploiting these observations, we propose an experimental scheme, named {\bf D}ump-produced {\bf A}boriginal {\bf M}atter {\bf S}earches at an {\bf A}ccelerator or DAMSA,\footnote{\begin{CJK}{UTF8}{}
In Korean characters \CJKfamily{mj}담사,
\end{CJK}\textipa{/d\textscripta\textlengthmark ms\textscripta/}: deep thought, rumination.} for the dark-sector investigation. 
For illustration purposes, we consider a RAON-like facility~\cite{Jeon:2014oja} and study the detection prospect of ALPs as a benchmark physics case for demonstrating capability of DAMSA
in the search for dark-sector physics and how the BRN can be evaded. 

Indeed, the low-energy, high-intensity, beam facilities are particularly suited for probing ALPs interacting with SM photons. 
First of all, a highly intensified beam can produce photons copiously in the beam dump through various production mechanisms including bremsstrahlung of secondary charged particles~\cite{Dutta:2020vop,Brdar:2020dpr}.
These photons can be converted to ALPs via the Primakoff effect, and thus an enormous flux of ALPs is generally expected. 
Second, since ALPs can decay into a pair of photons which are massless, the relevant experiment can probe even low-mass ALPs (say lower than twice the electron mass) in the decay channel, fully exploiting the photon flux as long as kinematics is allowed. 
Third, as discussed earlier, the expected (neutrino-induced) background rates are small, compared to those utilizing a higher-energy beam.  
Therefore, we expect that the DAMSA experiment at RAON-like facilities can provide great opportunities in terms of ALP searches. 

\begin{figure}[t]
    \centering
    \includegraphics[width=0.85\linewidth]{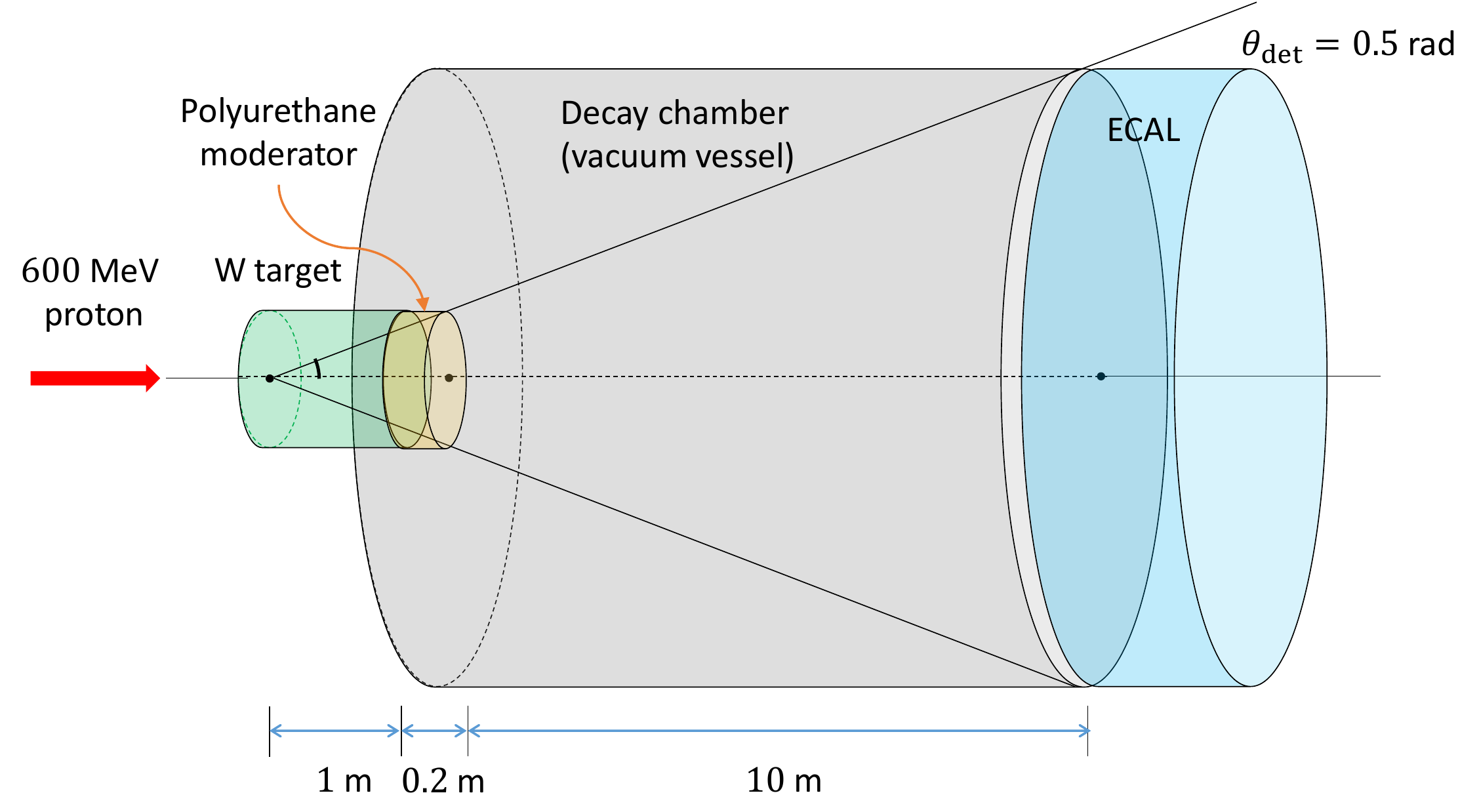}
    \caption{A conceptual layout of an example DAMSA experiment implemented at a RAON-like facility.
    }
    \label{fig:design}
\end{figure} 

\medskip

\noindent {\bf Experimental Concept.} 
We envision the situation that a 660~$\mu$A, 600-MeV proton beam provided by a RAON-like facility impinges on a 1-meter long cylindrical tungsten target, immediately followed by a polyurethane neutron moderator and a vacuum decay chamber whose volume is defined by a 10-meter long cylinder and whose wall is framed by 0.6-cm thick steel plates (see Fig.~\ref{fig:design}).
Electromagnetic particles are detected for identification at a high-granularity, high resolution ECAL located immediate downstream behind the decay chamber. 
We assume the ECAL angular coverage $\theta_{\rm det}$ to be as large as 0.5 radians to improve the ALP signal sensitivity, i.e., when a line emanating from the center of the top base of the target intersects the beam axis by $\theta_{\rm det}=0.5$~rad. 
We further assume that the energy threshold for photons $E_{\rm th}$ and the angular resolution $\theta_{\rm res}$ should be as low as 15 MeV and $1^\circ$, respectively, not only to maximize the ALP signal sensitivity but to achieve significant background rejection. 

Given the models of ALP interacting with photons, it is crucial to estimate spectral information of photons emerging in the target as precisely as possible for precise estimates of the ALP-signal sensitivity. 
To this end, we employ a dedicated detector-level simulation code package, \texttt{GEANT4}, with the QGSP\_BIC\_AllHP library~\cite{Agostinelli:2002hh}.
We deliver $10^6$ 600-MeV protons to the target and then find $\sim1.9\times 10^{8}$ photons inside the target.
For simplicity, we considered the target geometry and no other components which would be part of the target complex, expecting that such details would make an insignificant impact on our results.

\begin{figure}[t]
    \centering
    \includegraphics[width=8.4cm]{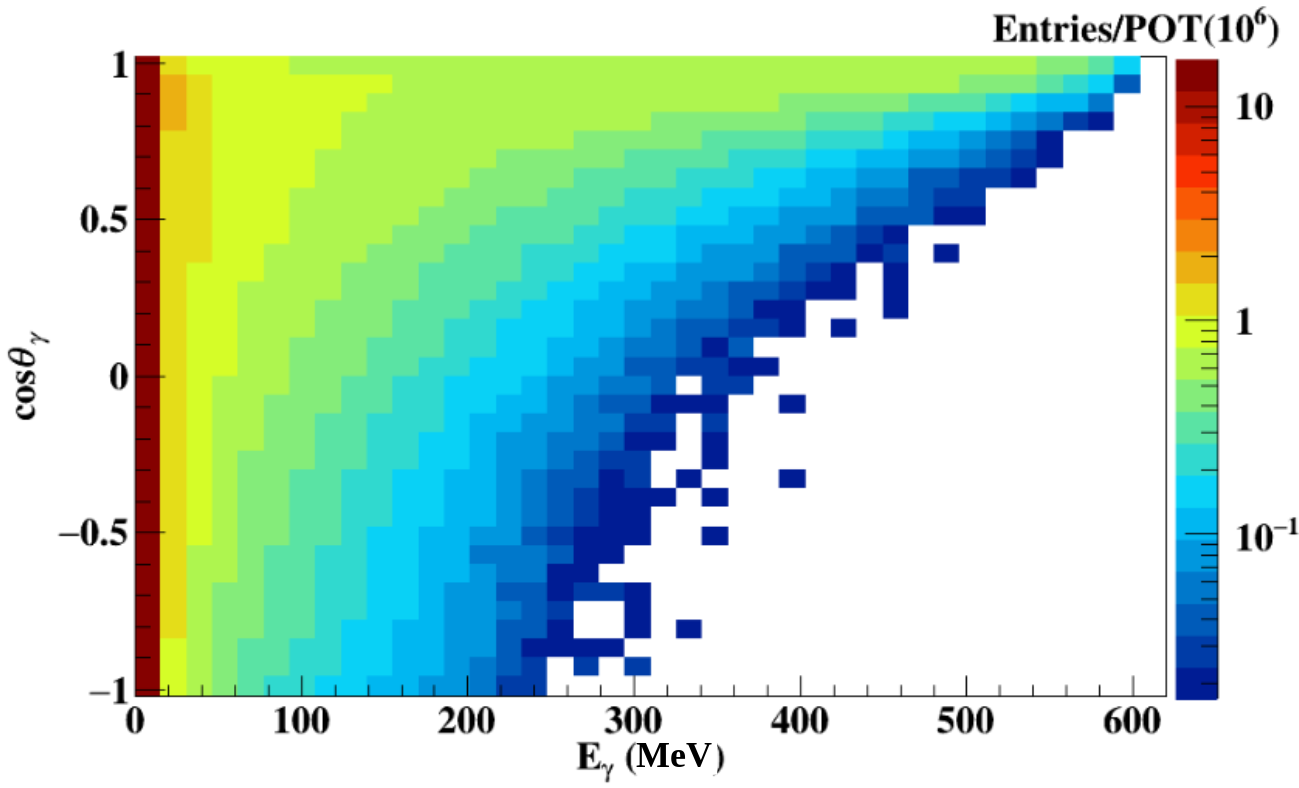}
    \caption{Two-dimensional correlation between energy and $\cos\theta_\gamma$ of photons produced in the target of the proposed experiment at a RAON-like facility.
    The angle is measured with respect to the beam axis.}
    \label{fig:photonspec}
\end{figure}

We display the correlation between the energy and the $\cos\theta_\gamma$ of photons produced in the target in Fig.~\ref{fig:photonspec}. 
First, soft photons dominate the energy spectrum. Nevertheless, a non-negligible fraction of photons with energy greater than 100 MeV can be produced, allowing DAMSA to be sensitive to $\mathcal{O}(100)$~MeV ALPs and set the new limits for them.
Second, the angular spectrum suggests that such energetic photons move preferentially in the forward direction.
Given the fact that our detector system covers up to $\theta_{\rm det}=0.5$~rad in the forward region, we find that about 7.5\% of photons will move toward the detector system. 

\medskip

\noindent {\bf ALP Signal.} 
We consider models of ALPs interacting with the SM photon for which the Lagrangian contains $\mathcal{L}_{\rm int} \supset - g_{a\gamma\gamma} \, a F_{\mu\nu} \Tilde{F}^{\mu\nu}/4$ where $g_{a\gamma\gamma}$ parameterizes the effective coupling strength between the ALP and the SM photon and $F_{\mu \nu}$ ($\tilde F_{\mu \nu}$) denotes the (dual) field strength tensor of the SM photon.
The presence of the above interaction allows a photon to convert to an ALP via the Primakoff process~\cite{Primakoff:1951iae}, $\gamma + A \to a +A$ with $A$ being the atomic system of interest.
Once produced, an ALP flies to the detector and can decay into two photons inside the decay chamber, $a\to 2\gamma$.
The decay width of ALP to a photon pair is well known and expressed as $\Gamma_a = g_{a\gamma\gamma}^2 m_a^3/(64\pi)$ with $m_a$ being the ALP mass.
Obviously, an ALP decays rather promptly if $m_a$ and/or $g_{a\gamma\gamma}$ are large.
Therefore, it is highly beneficial to place the detector as close to the target as possible to achieve improved sensitivities to decay signals of (relatively) short-lived ALPs.
In addition, an expected high flux of ALPs at DAMSA enables us to probe new regions of parameter space.

The data analysis strategy for the ALP signal of this sort is well known and standardized in literature (see, for example, Refs.~\cite{Dent:2019ueq,Brdar:2020dpr} and a review article \cite{Fortin:2021cog}). 
We follow the same strategy, modifying detector-specific parameters accordingly 
(see also Appendix~\ref{sec:app} for details).

\begin{figure}[t]
    \centering
    \includegraphics[width=0.49\linewidth]{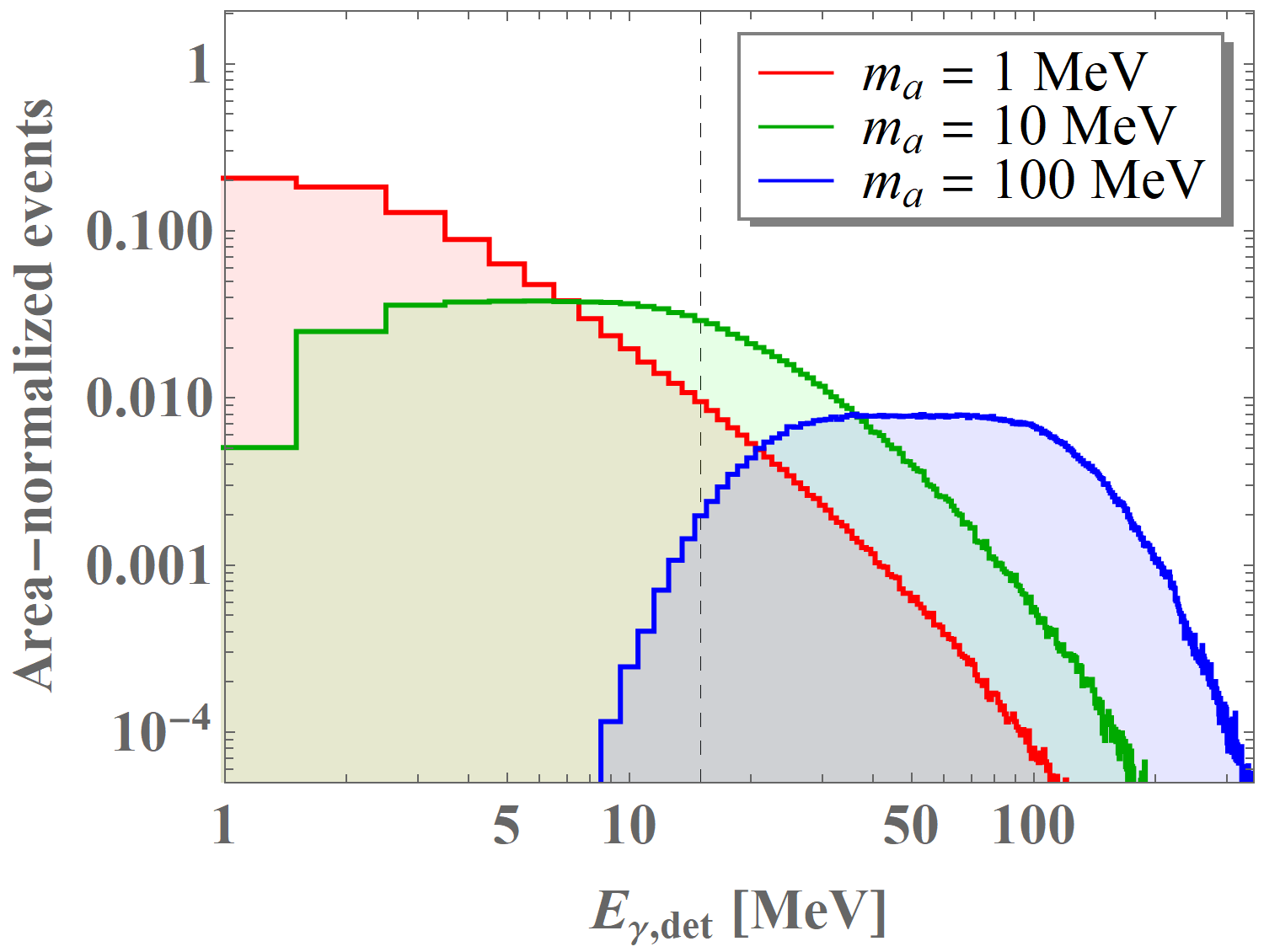} 
    \includegraphics[width=0.46\linewidth]{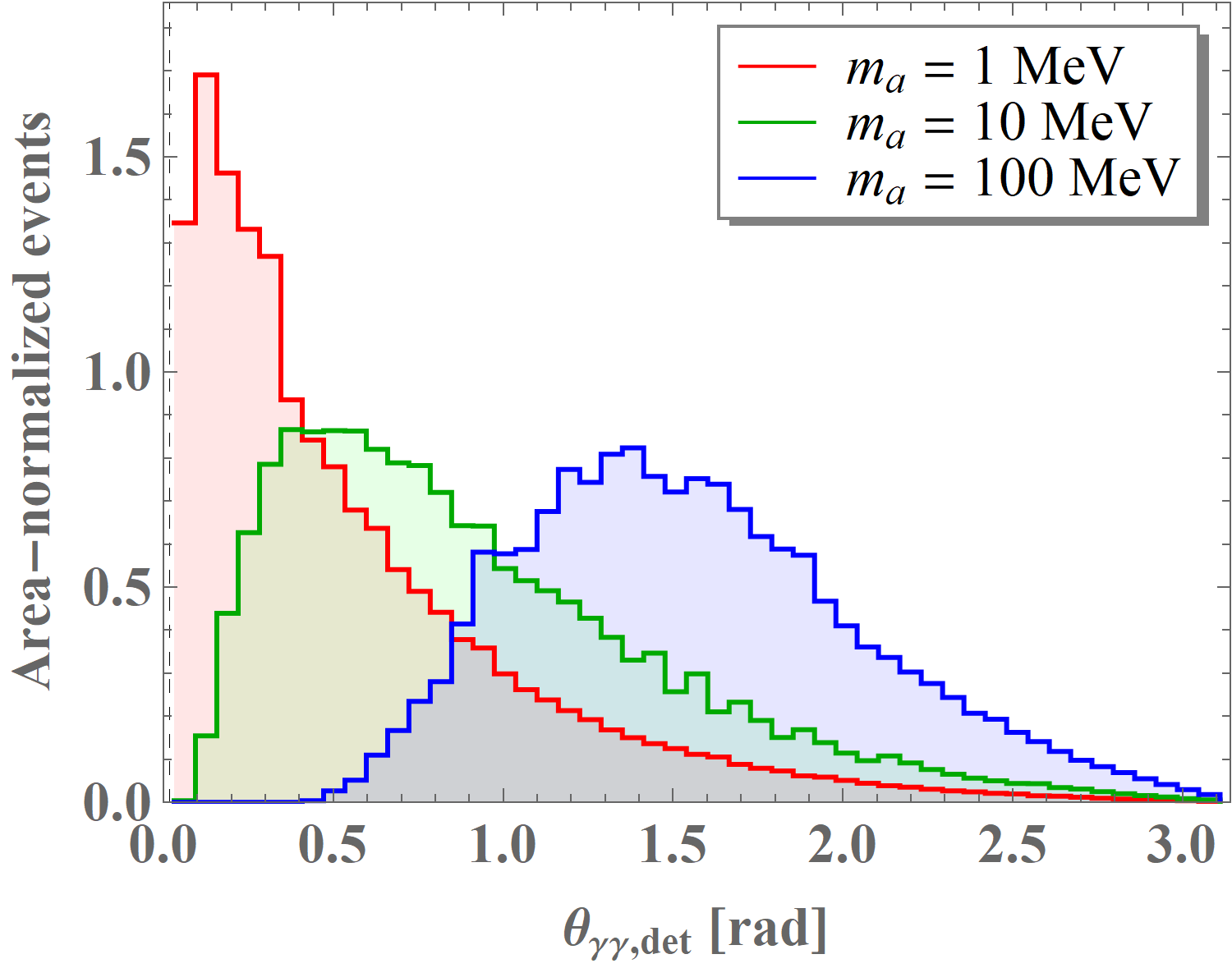}
    \caption{Left: Energy spectra of the photons from the ALP decay for three ALP mass values.
    The black dashed vertical line marks $E_{\rm th}=15$~MeV.
    Right: Spectra in the angular separation between the two final-state photons for the same three mass points.}
    \label{fig:kindist}
\end{figure}

Finally, we discuss kinematic distributions of ALP events at the detector, as they are related to the detector requirements and the cuts to discriminate the signal from potential backgrounds that we discuss shortly. 
Since the signal of our interest involves a pair of photons in the final state, we display spectra of $E_{\gamma,{\rm det}}$, the energy of individual photons, and spectra of $\theta_{\gamma\gamma,{\rm det}}$, the angular separation between the two photons in the left and right panels of Fig.~\ref{fig:kindist}. 

We remark that the energy spectra develop a log-symmetric structure with respect to the energy value being half the mass of ALP, since the ALP is a spinless particle and 
decays to two massless photons~\cite{Agashe:2012bn,Boddy:2016hbp}.
This implies that lower-mass ALP signals are more affected by the energy threshold or cut in their acceptance, since one of the two photons is less likely to pass the energy criterion. 
We show an example cut of 15~MeV by a dashed vertical line in the left panel of Fig.~\ref{fig:kindist}.
One can see that most of the photon pairs in the $m_a=100$~MeV case satisfy the cut, while large (some) fraction in the $m_a=1$ (10)~MeV case fails in passing the cut.  
In principle, softer energy cuts imposed on the final-state photons would be advantageous for increasing the signal acceptance in a wider range of parameter space, but the detailed choice also depends on the rejection capability of backgrounds. 

Regarding $\theta_{\gamma\gamma,{\rm det}}$, most of the two final-state photons are separated enough to be resolved with $\theta_{\rm res}=1^\circ$, as shown in the right panel of Fig.~\ref{fig:kindist}. 
Angular separations are likely to be larger with increasing $m_a$ since ALPs are less boosted. 
$\theta_{\gamma\gamma, {\rm det}}$ can be an additional handle to enhance the signal-over-background depending on $m_a$.

\medskip

\noindent {\bf Background Considerations.} 
We now discuss potential backgrounds to the ALP signal, considering non-neutron backgrounds followed by neutron-induced backgrounds and their rejection strategies. 

\noindent i) {\it \underline{Non-neutron backgrounds}}: 
Given the diphoton signature, possible SM backgrounds would be neutrino neutral-current single $\pi^0$ events occurring at the steel vacuum vessel, e.g., $\nu+{\rm Fe} \to \nu+\pi^0+{\rm Fe}$, since the $\pi^0$ decay gives rise to the same signature.
However, RAON-like facilities are essentially stopped-pion sources and the resulting neutrinos enter the detector with insufficient energy to form a $\pi^0$.
Therefore, we expect that this potential backgrounds will not arise at the proposed detector. 

The second class of backgrounds related to neutrinos is the scattering of neutrino off an electron in the vacuum vessel together with a hard photon emission, i.e., $\nu+e^- \to \nu+e^-+\gamma$, in which the final-state electron is misidentified as a photon.
Our simulation suggests that for $E_\nu > 15$~MeV, the scattering cross-section of the above process
with $E_{\gamma} > 15~{\rm MeV}$ is about $10^{-49}~{\rm cm}^2$ and $\sim 0.11$ neutrinos are produced per proton collision.
Considering a $\mathcal{O}(1)$-cm thickness of the vacuum vessel and the number density of target electrons of steel, $2.3\times 10^{24}$~cm$^{-3}$, we find that the number of this type of neutrino events created in a year is much less than 1. 

When it comes to non-neutrino-related backgrounds, particles coming out of the target end would induce signal-faking events.
Although the rate is expected to be small, the DAMSA detector system in immediate contact with the target may take a significant flux of such particles, making them non-negligible.
Given the beam energy, however, a 1-m long tungsten target also plays a role of shielding material.
Therefore, the (charged) particles will be almost completely absorbed before escaping from the target.
Our \texttt{GEANT4} simulation confirms this expectation and only $\mathcal{O}(2-3)$ photons and electrons may exit the target for $10^5$ proton collisions.
If any pairs of such photons and/or photon-faking electrons are accidentally reconstructed as if they came from the same vertex point inside the vacuum chamber within the timing and position resolutions of the ECAL, they will be mis-identified as signal events.
A similar type of potential background events can arise from neutron-induced events with a much higher rate, which will be discussed in the next part.

A similar issue would arise with cosmic particles because the detector will be essentially placed near the surface.
While beam triggering would greatly suppress such backgrounds, a comic tagger or active veto regions surrounding the detector fiducial volume may be needed.  

\smallskip

\noindent ii) {\it \underline{Neutron-induced backgrounds}}: 
The most important feature of the DAMSA concept is the proximity of the detector to the target, potentially resulting in a formidable level of neutron flux entering the detector system.
The experimental feasibility relies on the ability to distinguish the neutron-induced photons mimicking the signal.
We here propose a viable strategy to significantly reduce the neutron-induced backgrounds and discuss its performance. 

First of all, as mentioned earlier, we employ a moderator to suppress the number of neutrons potentially coming near the decay chamber or the detector. 
Our simulation with a 20-cm polyurethane moderator suggests that about 97\% of neutrons be shielded.\footnote{The shielding rate, of course, depends on the energy of the neutrons, and in general, low-energy neutrons can be effectively shielded by this method.} 
Second, we find that a 15-MeV energy threshold can reject about 99.99\% of photons induced by the remaining neutrons without deteriorating the signal acceptance.
Finally, we devise a few selection criteria applicable to any photon pairs, making use of the diphotonic nature of our ALP signal.

\begin{figure}[t]
    \centering
    \includegraphics[width=8.4cm]{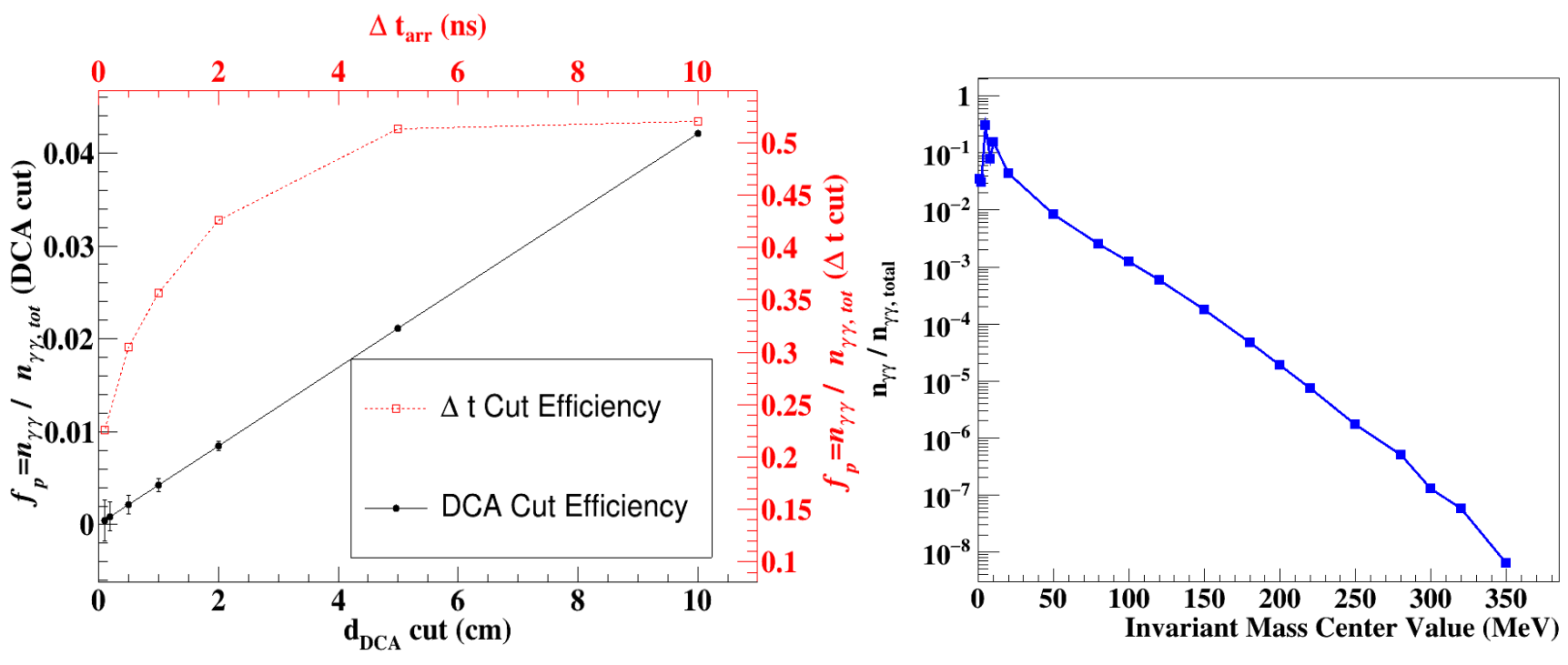}
    \caption{Left: Fractional number of photon pairs vs. $d_{\rm DCA}$ cut (black solid line) [$\Delta t_{\rm arr}$ cut (red dotted line)].
    Right: Fractional number of photon pairs against invariant mass. }
    \label{fig:DCAcut}
\end{figure}

\begin{itemize}
\item[C1.] {\it{Distance-of-closest-approach}} (DCA):
We introduce DCA ($d_{\rm DCA}$) which defines the closest distance between given two photon tracks.
If any two photons' extrapolated tracks cross over each other at a common point inside the decay chamber, such a photon pair is potentially mis-tagged as a signal.
DCA allows us to exclude the unrelated photon pairs from signal candidates.
The black line in the left panel of Fig.~\ref{fig:DCAcut} shows the fractional number of photon pairs in the $d_{\rm DCA}$ cut. 

\item[C2.] {\it{Arrival time difference}}: 
The temporal correlation of any two photons can be examined, as signal photons are produced simultaneously. 
We estimate the time difference ($\Delta t_{\rm arr}$) of two photons at the DCA position and discard any photon pairs showing a large time difference.
The red line in the left panel of Fig.~\ref{fig:DCAcut} shows the fractional number of photon pairs in the $\Delta t_{\rm arr}$ cut.

\item[C3.] {\it{Fiducial volume}}: 
Neutron-induced photons are produced mostly at the steel frame of the vacuum decay chamber, whereas signal photons emerge inside the decay volume.
We reject the photon pairs satisfying C1 and C2, if their DCA position is located inside the steel frame. 

\item[C4.] {\it{Trace-back analysis}}: 
The vectorial sum of the two photon three momenta, which would be considered as the ALP momentum, should be traced back to the target to be misidentified as a signal.
We extrapolate back the momentum vector from the DCA position and examine whether or not this extended line stems from the target.

\item[C5.] {\it{Invariant mass window}}: 
Finally, the reconstructed invariant mass of the two photons ($M_{\gamma\gamma}$) should agree with the $m_{a}$ value under consideration within detector mass resolution $\sigma_{m}$ for which we assume 1 MeV.
The right panel of Fig.~\ref{fig:DCAcut} shows the cut efficiency as a function of the invariant mass window. 
\end{itemize}

As above, five different independent selection criteria are used to mitigate the neutron-induced background, and their individual performance is summarized in Table~\ref{tab:cuttable}.

\begin{table}
\centering
\begin{tabular}{ c | c } 
\hline
\hline
Description (per pulse) & Number\\
\hline
Protons   & $4.8\times10^{7}$ \\ 
Beam-induced neutrons   & $1.29\times10^{5}$  \\ 
Neutron-induced photons   & $2.74\times10^{5}$  \\
Photons with $E_\gamma>15$~MeV    & $25.1$ \\
Photons hitting the detector & $2.94$  \\
$<40$ ns arrival time cut   & $1.47$  \\
Photon pairs  & $1.08$  \\
\hline
Yearly number of pulses & $2.93\times10^{15}$  \\
\hline \hline
Photon pair selection criteria  & Efficiency  \\
\hline
$d_{\rm DCA}< 1$ cm   & $4.23\times10^{-3}$     \\
$\Delta t_{\rm arr} < 0.1$ ns  &$ 2.01\times10^{-1}$    \\
Fiducial volume cut  & $6.13\times10^{-1}$    \\
Back-tracing  & $4.16\times10^{-2}$     \\
\hline
$M_{\gamma\gamma}\in 50\pm 1$~MeV & $8.34\times10^{-3}$\\
$M_{\gamma\gamma}\in 100\pm 1$~MeV & $1.25\times10^{-3}$\\
$M_{\gamma\gamma}\in 200\pm 1$~MeV & $1.91\times10^{-4}$\\
\hline \hline
\end{tabular}
\caption{A summary of the numbers related to the neutron-induced background mitigation and individual performance of the cuts. 
}
\label{tab:cuttable}
\end{table}

\medskip

\noindent {\bf Results.} 
In our analysis, we assume that $\sim 1.4\times 10^{23}$ protons are annually delivered to the target, and the results are reported for one-year and 10-year exposures.
Our \texttt{GEANT4} simulation shows that most of the photons are created at the position about 20 cm away from the facade of the target, so we assume that all of the photons emerge at this position for simplicity.
Furthermore, since a 20-cm long polyurethane moderator is placed between the target and the decay chamber to mitigate the neutron flux out of the target, $\ell$, the distance between the ALP production vertex and the decay chamber is $\sim 1$~m. 

However, depending on the incident angle of ALP, $\ell$ varies.
In our analysis, we set $\ell$ to be the average value, 1.04~m.
Likewise, the cord length that ALPs would travel along is about 10~m, and $L=10.4$~m is used as a baseline value.  
The production cross-section of ALPs is computed in the collinear limit, i.e., $E_\gamma\approx E_a$, as the momentum transfer to the target nucleus is much smaller than $E_\gamma$.
$\sigma_{\rm SM}$ is generally a function of $E_\gamma$~\cite{xcom}, and we consider photons of $E_\gamma>15$~MeV.  
If the photon energy is large enough, the change in $\sigma_{\rm SM}$ is not large. However, we will utilize $\sigma_{\rm SM}=\sigma_{\rm SM}(E_\gamma)$ for a more precise estimate, according to the measurement data archived in \cite{xcom}.
Finally, we envision a detector with $E_{\rm th}=15~{\rm MeV}$ and $\theta_{\rm res}=1^\circ$ as mentioned earlier, and require two final-state photons satisfying the following baseline criteria:
\begin{itemize}
    \item their reconstructed tracks intersect at a point inside the decay chamber within the timing and position resolutions of ECAL, $t_{\rm res}=0.1$~ns and $\vec{x}_{\rm res}=1$~cm,
    \item the diphoton invariant mass is consistent with $m_a$ under consideration within $\pm 1$~MeV window, and
    \item their vectorial 3-momentum sum, which is considered to be the same as the ALP momentum, traces back toward the target and the vertex position should be located inside the fiducial volume of the decay chamber.  
\end{itemize}
As we discussed in the previous section, these criteria allow us to suppress backgrounds including the neutron-induced by more than 7 orders of magnitude.
As conventional electromagnetic calorimetric systems can attain the above-described level of detector performance, we expect that the resultant sensitivity reaches can be considered rather conservative. 

\begin{figure}[t]
    \centering
    \includegraphics[width=0.99\linewidth]{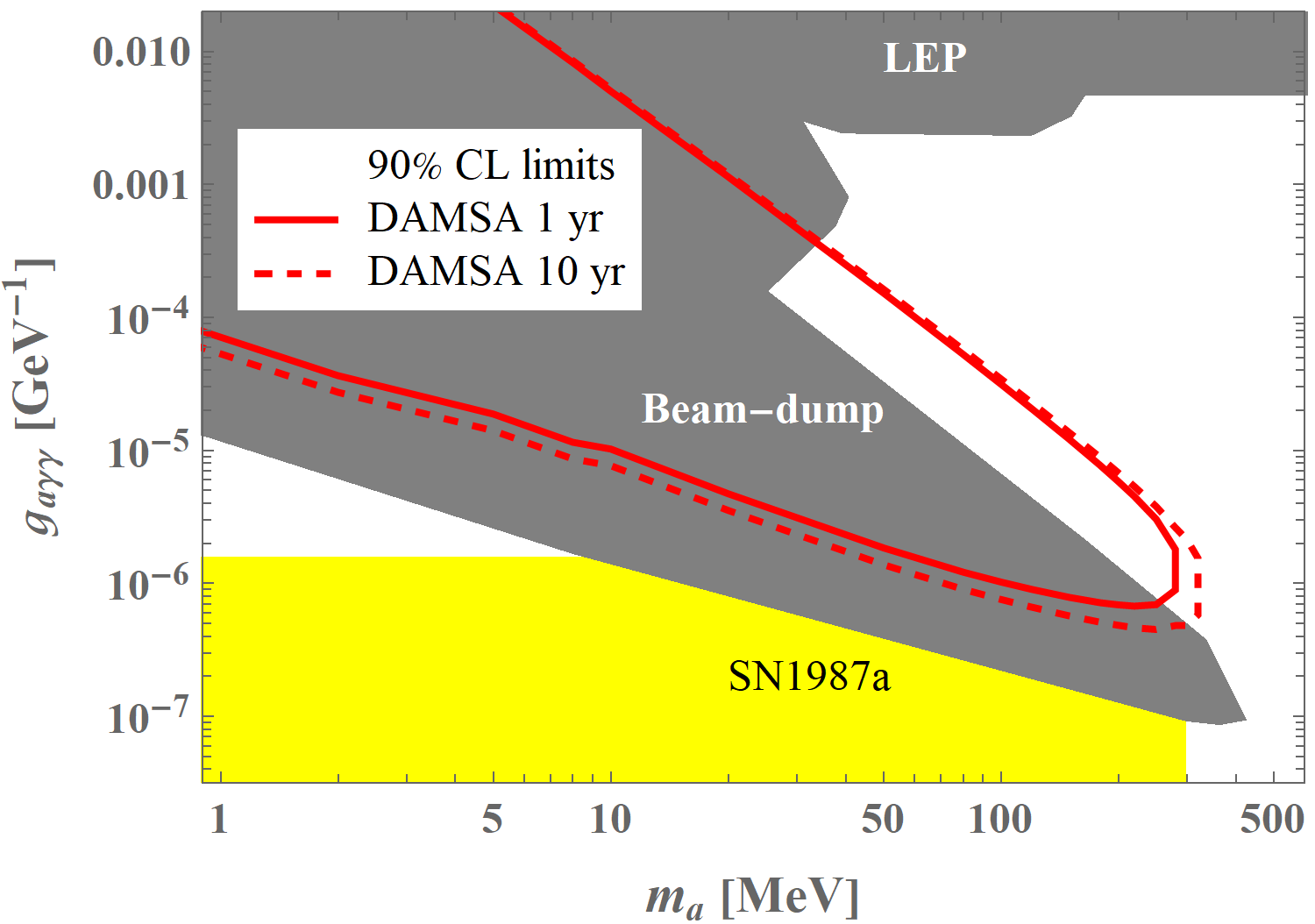}
    \caption{Expected 90\% C.L. sensitivity reaches of the ALP decay signal in the DAMSA experiment at a RAON-like facility (based on expected statistical error only) with 1-year (solid) and 10-year (dotted) exposures, under the assumption of $1.4\times 10^{23}$ POTs per year. 
    }
    \label{fig:result}
\end{figure}

We now report our results of the 90\% C.L. sensitivity reaches expected at DAMSA, in the $(m_a, g_{a\gamma\gamma})$ plane in Fig.~\ref{fig:result}. 
The limits are evaluated in the basis of expected statistical error only, for both one-year (solid) and 10-year (dotted) exposures.
We also show the existing excluded regions based on the limits compiled in e.g., Refs.~\cite{Bauer:2018uxu,Lanfranchi:2020crw}.
The gray-colored covers the regions excluded by the laboratory-produced ALP searches, and the boundaries are set by various lepton collider and beam-dump experiments. 
By contrast, the regions constrained by astrophysical ALP searches (e.g., HB stars, supernova, etc) are shown in yellow.

Our findings suggest that DAMSA can explore the regions beyond the current beam-dump limits, i.e., $m_a\sim 100$~MeV in-between $g_{a\gamma\gamma}\sim 10^{-6}~{\rm GeV}^{-1}$ and $\sim 10^{-4}~{\rm GeV}^{-1}$, which none of the existing (in)direct experiments have ever probed.
The somewhat limited beam energy relative to those in ongoing and upcoming beam-dump type experiments would not allow for copious production of $\mathcal{O}(100)$~MeV ALPs. However, close proximity of the DAMSA detector system to the target can offset such lack of statistics by capturing a sizable fraction of rather promptly decaying such ALPs.

\medskip

\noindent {\bf Conclusions.} 
We have proposed an experiment that would exploit the high-intensity proton beams at rare isotope nuclear physics facilities to search for ALPs, envisioning a facility similar to RAON in South Korea for illustration.  
The results show a great potential to cover a wide range of unexplored parameter space even with one year of data taking, thanks not only to the high flux proton beams but also to the lower proton energy and the proximity of the DAMSA detector system to the dump.
These advantages could also open a great potential for expanding physics opportunities beyond the ALP.  

\acknowledgements

We thank Yeonsei Chung and Taeksoo Shin in the RAON Construction Team at the Institute for Basic Science, Deuksoon Ahn at Korea Basic Science Institute, and Hyunjun Shin at POSTech for insightful discussions.
The work of WYJ and JY is supported by the U.S. Department of Energy (DOE), HEP Award No. DE-SC0011686.
The work of DK is supported by DOE under Grant No. DE-FG02-13ER41976, DE-SC0009913, DE-SC0010813. 
The work of KK is supported by the DOE No. DE-SC0021447. 
The work is supported by the National Research Foundation of Korea (NRF) [NRF-2021R1A4A2001897 (JCP, YK), NRF-2019R1C1C1005073 (JCP), NRF-2018R1A6A1A06024977/NRF-2020K1A3A2A01084612/NRF-2018R1A6A1A06024970/NRF-2019R1A2C1008544 (MSR), and NRF-2020R1I1A3072747/NRF-2022R1A4A5030362 (SS)].

\appendix
\section{Data Analysis \label{sec:app}}

\begin{figure}[t]
    \centering
    \includegraphics[width=7.4cm]{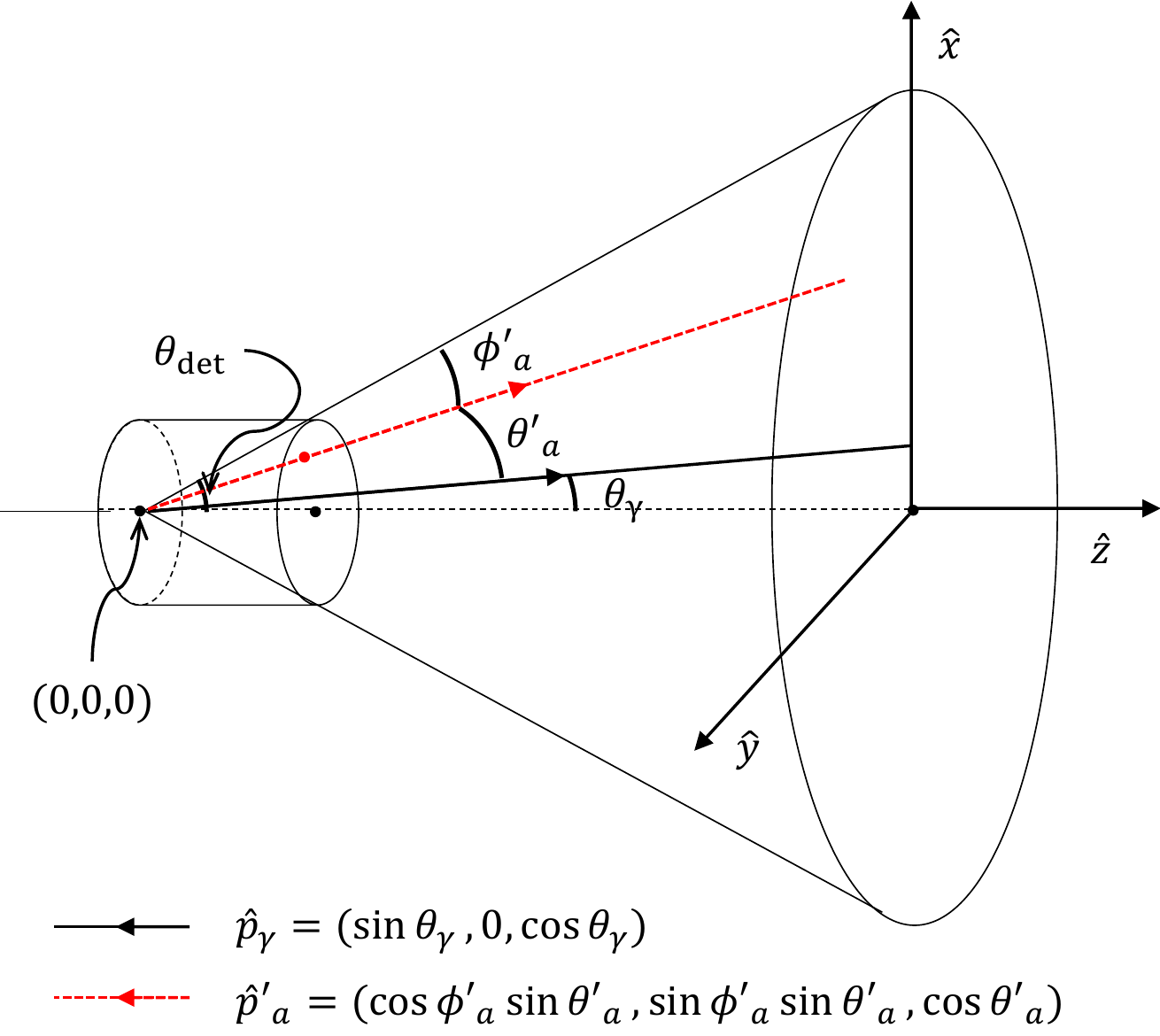}
    \caption{An example configuration of an incoming photon (black solid arrow coming from the origin), an outgoing ALP (red dashed arrow coming from the origin), and a detector.
    The photon is assumed to emerge and convert to an ALP at $(0,0,0)$.
    The unprimed angles are defined with respect to $\hat{z}$ (i.e., the beam axis), while the primed angles are defined with respect to $\hat{p}_\gamma$ (i.e., the momentum direction of the initial-state photon). 
    }
    \label{fig:coord}
\end{figure}

We focus on the calculation of the expected signal rate at the detector for a given single photon of energy $E_\gamma$.
The production cross-section $\sigma_P$ as a function of the ALP angular variables is given by
\begin{equation}
    \frac{d^2\sigma_P}{d\theta'_a d\phi'_a}= \frac{1}{8\pi} g_{a\gamma\gamma}^2\alpha Z^2 \left[F(t)\right]^2\frac{p_a^4\sin^3\theta'_a}{t^2}\,, \label{eq:primakoff}
\end{equation}
where $\alpha$, $Z$, and $p_a$ are the electromagnetic fine structure constant, the atomic number of the target material, and the momentum of the outgoing ALP, and momentum transfer $t=m_a^2-2E_\gamma(E_a-p_a\cos\theta'_a)$.
In the collinear limit where the momentum transfer to the atomic system is negligible, $E_a\approx E_\gamma$, and we work on in this limit throughout this paper.
Here $\theta'_a$ and $\phi'_a$ are the polar and the azimuthal angles of ALP measured from the incoming photon direction.
See also Fig.~\ref{fig:coord} for more detailed configuration. 
Finally, $F$ describes a form factor for which we take the Helm parameterization,
\begin{equation}
    F(t)=\frac{3j_1(\sqrt{|t|}R_1)}{\sqrt{|t|}R_1}\exp\left(-\frac{|t|s^2}{2} \right)\,,
\end{equation}
where $j_1$ denotes the spherical Bessel function.
Here $s=0.9$~fm and $R_1=\sqrt{(1.23A^{1/3}-0.6)^2+2.18}$~fm with $A$ being the atomic mass number of the target material as per Ref.~\cite{Lewin:1995rx}.

Inside the target material, the Primakoff process in \eqref{eq:primakoff} competes with the SM processes such as pair production and photoelectric absorption that the photons usually go through.
Therefore, the probability of ALP production, say $P_{\rm prod}$, is given by the ratio of the total ALP production cross-section to the total cross-section of SM interactions $\sigma_{\rm SM}$:
\begin{equation}
    P_{\rm prod}=\frac{1}{\sigma_{\rm SM}+\sigma_P}\int d\theta'_a d\phi'_a \frac{d^2\sigma_P}{d\theta'_a d\phi'_a}\,.
\end{equation}

To determine the integration limits, we consider the coordinates and angles defined in Fig.~\ref{fig:coord}.
Assuming that a photon emerging at the origin intersects with the beam axis by an angle of $\theta_\gamma$, we see that the unit vector of the photon direction $\hat{p}_\gamma$ is given by $\hat{p}_\gamma=(\sin\theta_\gamma,0,\cos\theta_\gamma)$.
Then $\theta'_a$ and $\phi'_a$ mentioned above define the unit vector of the outgoing ALP $\hat{p}'_a$ about $\hat{p}_\gamma$: $\hat{p}'_a=(\cos\phi'_a\sin\theta'_a,\sin\phi'_a\sin\theta'_a,\cos\theta'_a)$.
A simple algebra enables us to find that the polar angle of ALP $\theta_a$ measured from the beam axis is
\begin{equation}
    \theta_a = \cos^{-1} (\cos\theta_a^\prime \cos\theta_\gamma + \cos\phi_a^\prime \sin\theta_a^\prime \sin\theta_\gamma)\,.
    \label{eq:angle}
\end{equation}
If $\theta_a \leq \theta_{\rm det}$, then the associated ALP can enter the detector.
One can find that the maximum and minimum $\theta'_a$ values obeying this requirement are 
\begin{eqnarray}
    \theta'_{a,+}&=&\min(\pi, \theta_\gamma+\theta_{\rm det})\,, \\
    \theta'_{a,-}&=&\min\left[\max(0,\theta_\gamma-\theta_{\rm det}),2\pi-\theta_\gamma-\theta_{\rm det} \right]\,.
\end{eqnarray}
When it comes to the limits of $\phi'_a$, we again rely on the $\theta_a \leq \theta_{\rm det}$ condition from which we obtain
\begin{equation}
    \phi'_{a,\pm} =\pm \cos^{-1}\left( \max\left[-1,\frac{\cos\theta_{\rm det}-\cos\theta'_a\cos\theta_\gamma}{\sin\theta'_a\sin\theta_\gamma}\right] \right)\,.
\end{equation}

\begin{figure}[t]
    \centering
   \includegraphics[width=7.4cm]{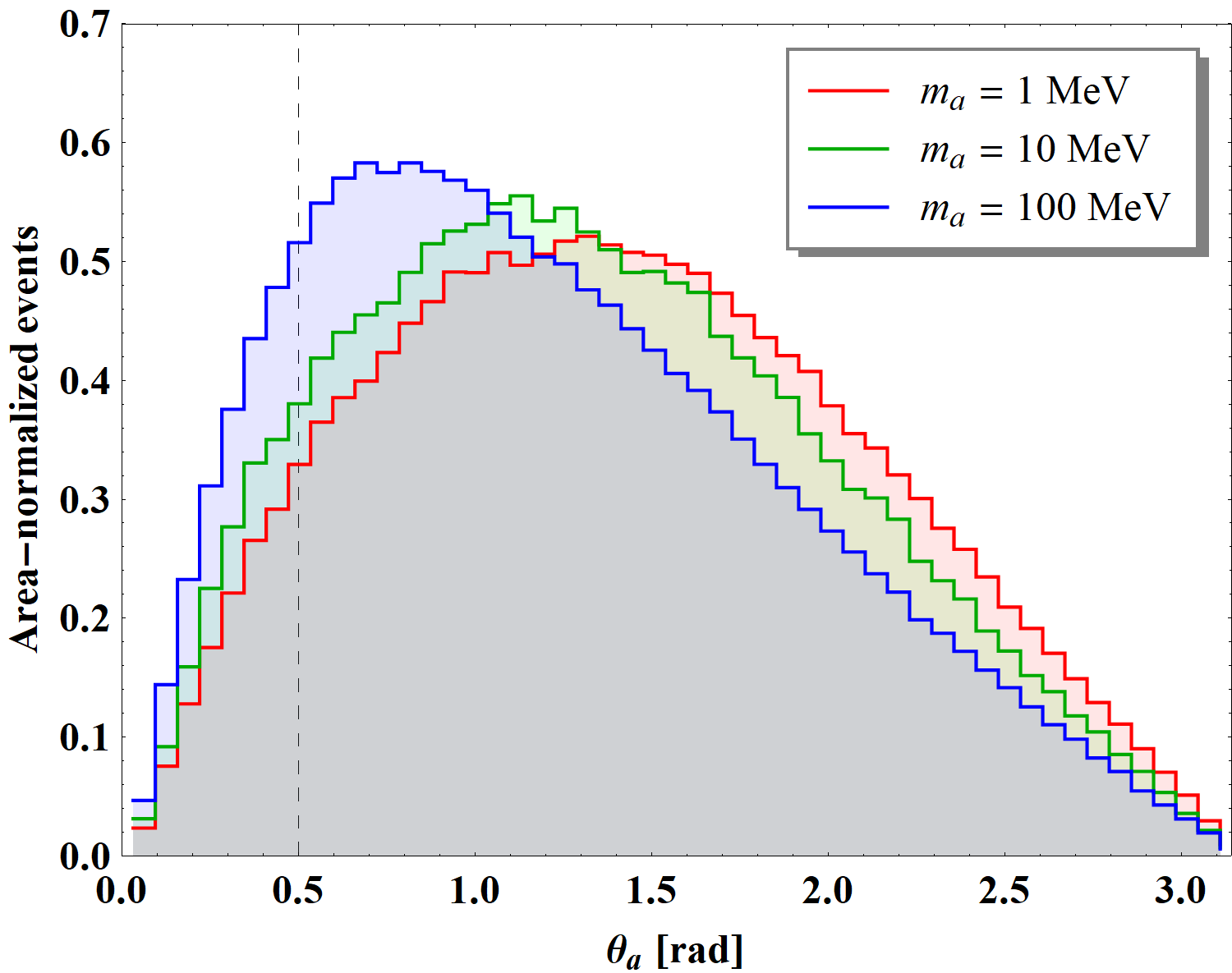}
    \caption{Angular spectra of ALPs produced in the target for three representative masses: $m_a=1$~MeV, 10~MeV, and 100~MeV.
    The black dashed vertical line indicates $\theta_a=0.5$~rad by which the coverage of the proposed detector will be defined. }
    \label{fig:ALPangle}
\end{figure}

Before moving onto the next step, it is instructive to see the (area-normalized)
ALP spectra in $\theta_a$ predicted by the formalism thus far in combination with the photon flux generated, since they allow for quantitative estimates on what fraction of the produced ALP flux would traverse the detector.
We show the expected angular spectra in Fig.~\ref{fig:ALPangle} for three representative mass values: $m_a=1$~MeV (red), 10~MeV (green), and 100~MeV (blue).
For reference purposes, we add a black, dashed, vertical line at $\theta_a=0.5$~rad below which the ALPs would enter the detector.
We see that the spectrum for $m_a=1$~MeV does not show a substantial difference from the corresponding photon angular spectrum, but those for larger $m_a$ are more forward-directed.
The reason is as follows.
Basically, most of the ALPs are created in the forward region with respect to $\hat{p}_\gamma$, as also suggested in \eqref{eq:primakoff}.
Energetic photons responsible for production of heavy ALPs, however, are populated relatively more along the beam direction, while soft photons are more broadly distributed.
We find that about 10\% of the ALP flux (i.e., below the black, dashed, vertical line) would move toward the detector for $m_a=100$~MeV.  

Once a produced ALP lies in a right direction, it should survive, not decay before reaching the facade of the detector fiducial region. 
It is simply described by the usual decay law.
Suppose that the distance between the ALP production vertex and the detector is $\ell$.
The survival probability $P_{\rm surv}$ is then given by
\begin{equation}
        P_{\rm surv}=\exp\left(-\frac{\ell}{\ell_a^{\rm lab}}\right)\,,
\end{equation}
where $\ell_a^{\rm lab}$ denotes the laboratory-frame mean decay length of ALP which is a well-known function of the decay width and the boost factor of ALP $\gamma_a$:
\begin{equation}
    \ell_a^{\rm lab}=\frac{\sqrt{\gamma_a^2-1}}{\Gamma_a}\,.
\end{equation}

On the other hand, an ALP reaching the detector should decay to a photon pair before escaping from the detector fiducial region to leave an experimental signature.
It is again described by the decay law.
Assuming that the cord length along which the ALP would sweep is denoted by $L$, we write the decay probability $P_{\rm dec}$ as follows:
\begin{equation}
    P_{\rm dec}= 1-\exp\left(-\frac{L}{\ell_a^{\rm lab}} \right)\,.
\end{equation}

Collecting all the probabilities together, we find that the expected number of ALP events for any $i$th injection photon (denoted by $N_{a,i}$) can be written as
\begin{equation}
    N_{a,i}=P_{\rm prod}\times P_{\rm surv} \times P_{\rm dec}\times 1\,. \label{eq:Nai}
\end{equation}
Note here that the last factor ``1'' in the right-hand side is added explicitly since we consider a {\it single} injection photon.  
Therefore, if $N_\gamma$ photons are produced in the target for a given exposure, the total number of ALP events $N_{a,{\rm tot}}$ equals to the summation over $N_\gamma$:
\begin{equation}
    N_{a,{\rm tot}}=\sum_{i=1}^{N_\gamma}N_{a,i}\,. \label{eq:Ntot}
\end{equation}
One can calculate an average of $N_{a,i}$ in \eqref{eq:Nai} over a sufficiently large sample of photons, say $\langle N_{a,i} \rangle$.
This can be interpreted as acceptance of the ALP signal, and $N_{a,{\rm tot}}$ in \eqref{eq:Ntot} can be calculated alternatively by 
\begin{equation}
    N_{a,{\rm tot}}= N_\gamma \langle N_{a,i} \rangle = N_{\rm POT} R_\gamma \langle N_{a,i} \rangle\,,
\end{equation}
where in the second equality we factorize $N_\gamma$ into POT for a given exposure $N_{\rm POT}$ and the average production rate of photons per incident proton $R_\gamma$. 

\bibliography{ref}
\bibliographystyle{JHEP}

\end{document}